\documentstyle[12pt,epsf]{article}
\begin{document}
\baselineskip 16pt plus 2pt minus 2pt

{\bf VERSION \today}

\vspace{0.3cm}

\begin{center}

{\large \bf   THE REACTION \boldmath{$\pi N \to \pi \pi N$} AT THRESHOLD}

\vspace{2.5cm}

V. Bernard$^{\dag ,1}$, N. Kaiser$^{\S ,2}$, Ulf-G. Mei\ss ner$^{\dag ,3}$

\vspace{0.5cm}

$^{\dag}$Physique Th\'{e}orique, Centre de Recherches Nucl\'{e}aires et
Universit\'{e}
Louis Pasteur de Strasbourg, B.P. 20, F-67037 Strasbourg Cedex 2, France

\vspace{0.5cm}

$^{\S}$Technische  Universit\"{a}t M\"{u}nchen, Physik Department T30,
James-Franck-Stra\ss e, D-85747 Garching, Germany

\vspace{0.5cm}

email: $^1$bernard@crnhp4.in2p3.fr, $^2$nkaiser@physik.tu-muenchen.de,
$^3$meissner@crnvax.in2p3.fr

\end{center}

\vspace{3cm}

\begin{center}

{\bf ABSTRACT:}

\end{center}

\vspace{0.1cm}

We consider the chiral expansion for the reaction $\pi N \to \pi \pi N$ in
heavy baryon chiral perturbation theory. To order $M_\pi$ we derive novel
low--energy theorems that compare favorably with recent determinations of the
total cross sections  for $\pi^+ p \to \pi^+ \pi^+ n$ and
$\pi^- p \to \pi^0 \pi^0 n$.

\vspace{5cm}

\noindent CRN-94/19 \hfill April 1994

\vfill

\pagebreak

\vspace{0.5cm}

\noindent {\bf 1.} Over the last few years, new data for the reaction $\pi N
\to \pi \pi N$ in the threshold region and above have become available, see
e.g. \cite{poc} \cite{bulo} \cite{sevi} \cite{kern} \cite{lowe} and the
compilation in \cite{beri}. The interest in this reaction stems mostly from the
fact that it apparently offers a possibility of determining the low--energy
$\pi
\pi$ elastic scattering amplitude whose precise knowledge allows to test our
understanding of the chiral symmetry breaking of QCD. However, at present no
calculation based on chiral perturbation theory  is available which links the
pion production data to the $\pi \pi \to \pi \pi$ amplitude in a {\it
model--independent} fashion. Consequently, all presently available
determinations of the S--wave $\pi \pi$ scattering lengths from the
abovementioned data should be taken {\it cum grano salis}. In the framework of
relativistic baryon chiral perturbation theory, Beringer \cite{beri} has
considered tree diagrams. In that approach, however, there is no strict
one--to--one correspondence between the loop and the small momentum expansion
due to the nonvanishing nucleon mass in the chiral limit \cite{GSS}. This
problem can be circumvented if one considers the nucleons in the
 non--relativistic limit. This was first used by Gasser and Leutwyler
\cite{GLR} (and others) and later formulated in
terms of heavy quark effective field theory methods in ref.\cite{jm}.
We will make use here of the
two--flavor formulation detailed in ref.\cite{BKKM}.
The aim of this letter is to
show that the first two terms in the chiral expansion of the threshold
amplitudes for $\pi N \to \pi \pi N$ lead to a set of low--energy theorems
which indeed can be tested against the available data for $\pi^+ p \to \pi^+
\pi^+ n$ and $\pi^- p \to \pi^0 \pi^0 n$ close to threshold.\footnote{This
result was indirectly contained in ref.\cite{beri} but
not made explicit and is much more transparent
in the formulation used here.} Naturally, at the next stage one has to
consider the following terms in the chiral expansion to make contact with the
$\pi \pi$ interaction.

\vspace{0.3cm}

\noindent {\bf 2.} To be specific,
 consider the process $\pi^a N \to \pi^b \pi^c N$,
with $N$ denoting the nucleon (proton or neutron) and '$a,b,c$' are isospin
indices. At threshold, the transition matrix--element
in the $\pi^a N$ centre--of--mass frame takes the form

\begin{equation}
T \, = \, i \, {\vec \sigma} \cdot {\vec k} \left[ D_1 (\, \tau^b \delta^{ac} +
\, \tau^c \delta^{ab} ) \, + \, D_2 \, \tau^a \delta^{bc} \right]
\label{e1}
\end{equation}

\noindent where $\vec k$ denotes the three--momentum of the incoming pion and
the amplitudes $D_1$ and $D_2$
 will be subject to the chiral expansion as discussed below. They are related
to the more commonly used amplitudes ${\cal A}_{2I,I_{\pi \pi}}$, with $I$ the
total isospin of the initial $\pi N$ system
and $I_{\pi \pi}$ the isospin of the two--pion system in the
final state, via

\begin{equation}
{\cal A}_{32} \, = \, 2\sqrt{2} \, D_1, \quad {\cal A}_{10} \, = \, -2D_1 \,
- \, 3D_2
\label{e2}
\end{equation}

\noindent which have recently been determined \cite{bulo}. In what follows,
we will also consider the total cross section for the reactions
$\pi^+ p \to \pi^+ \pi^+ n$ and $\pi^- p \to \pi^0 \pi^0 n$. At present,
only in these two channels there exist accurate data in the $20...30$ MeV
region above threshold. The data of ref.\cite{poc} for $\pi^+ p \to \pi^+ \pi^0
p$ are still too sparse and inaccurate in the threshold region which we are
investigating. Assuming that
the amplitude in the threshold region can be approximated by the exact
threshold amplitude, the total cross section can be written in a compact
form,

\begin{equation}
\sigma_{\rm tot}(s) =
 \frac{m^2}{2s} \, \sqrt{\lambda(s,m^2,M_\pi^2)} \, \Gamma_3
(s) |\eta_1 D_1 \, + \, \eta_2 D_2 |^2 \, S
\label{e3}
\end{equation}

\noindent with $m$ the nucleon and $M_\pi$ the pion mass, respectively
 and $s$ the total
centre--of--mass energy squared. $\Gamma_3 (s)$ denotes the conventional
three--body phase space and $\lambda(x,y,z)$ the K\"all{\'e}n--function. The
$\eta_{1,2}$ are channel-dependent isospin factors and $S$ is a Bose
symmetry factor.
For $\pi^+ p \to \pi^+ \pi^+ n$ and $\pi^- p \to \pi^0 \pi^0 n$ we have
$\eta_1 = 2 \sqrt{2}, \, \eta_2 = 0, \, S = 1/2$ and
$\eta_1 = 0 , \, \eta_2 = \sqrt{2}, \, S = 1/2$, in order. In the threshold
region, one can approximate to  a high degree of accuracy the three--body phase
space and flux factor \cite{bkm} by analytic expressions so that

\begin{equation}
\sigma_{\rm tot} (T_\pi) = \frac{M_\pi^2 \sqrt{3(2+\mu)(2+3\mu)}}{128 \pi^2
(1+2\mu )^{11/2} }  |\eta_1 D_1 \, + \, \eta_2 D_2 |^2 \, S \,
(T_\pi - T_\pi^{\rm thr})^2
\label{e4}
\end{equation}

\noindent Here, $T_\pi$ is the pion kinetic energy in the laboratory frame,
 $T_\pi = (s - m^2_1 - M_{\pi 1}^2) / (2 m_1) - M_{\pi 1}$ where the subscript
$'1'$ denotes the particles in the initial state. We furthermore have
introduced the small parameter $\mu =  M_\pi / m \simeq 1/7$.
This completes the necessary formalism.

\noindent{\bf 3.} In QCD, the chiral expansion of the amplitude functions
$D_1$ and $D_2$ takes the form\footnote{Here, $D$ stands as a generic
symbol for $D_{1,2}$.}

\begin{equation}
D = f_0 \, + \, f_1 \, \mu \, + \ f_2 \, \mu^2 \, + \ldots
\label{e5}
\end{equation}

\noindent modulo logarithms.
We are interested here in the first two coefficients of this
expansion. To calculate them, we make use of heavy baryon chiral perturbation
theory as detailed in ref.\cite{BKKM}. The pertinent effective
Lagrangian has the form

\begin{equation}
{\cal L}_{\rm eff} = {\cal L}_{\pi N}^{(1)} + {\cal L}_{\pi N}^{(2)}
+ {\cal L}_{\pi \pi}^{(2)}
\label{e6}
\end{equation}
with ${\cal L}_{\pi N}^{(1,2)}$ given in ref.\cite{BKKM} and the standard
meson Lagrangian e.g. in ref.\cite{GL}. The diagrams with insertions from
${\cal L}_{\pi N}^{(1,2)}$ which are non--vanishing at threshold are shown in
fig.1. Notice that the much debated next--to--leading order $\pi \pi$
interaction does not appear at this order in the chiral expansion.
\noindent It is important
to note that from ${\cal L}_{\pi N}^{(2)}$ only terms which
are kinematical $1/m$ corrections contribute. None of the low--energy
constants $c_{1,2,3}$ related to elastic $\pi N$ scattering (in particular to
the isospin--even S--wave scattering length $a^+$) \cite{pinc}
appear at order $q^2$ (here, $q$ denotes a small momentum
or a meson mass).  One can therefore write
down low--energy theorems for $D_{1,2}$
which only involve well--known physical (lowest order) parameters,
\begin{equation}
D_1  = \frac{g_A}{8 F_\pi^3} \left( 1 + \frac{7 M_\pi}{2 m} \right)
 + {\cal O}(M_\pi^2) \label{e7}
\end{equation}

\begin{equation}
D_2  = -\frac{g_A}{8 F_\pi^3} \left( 3 + \frac{17 M_\pi}{2 m} \right)
 + {\cal O}(M_\pi^2) \label{e8}
\end{equation}

\noindent with $g_A$ the axial--vector coupling constant. In what follows,
we will always use the Goldberger--Treiman relation $g_A = g_{\pi N} F_\pi / m$
to calculate the numerical values of $D_{1,2}$ (with $g_{\pi N}$ the strong
pion--nucleon coupling constant).

\vspace{0.5cm}

\hspace{0.5in}
\epsfysize=1.5in
\epsfxsize=5in
\epsffile{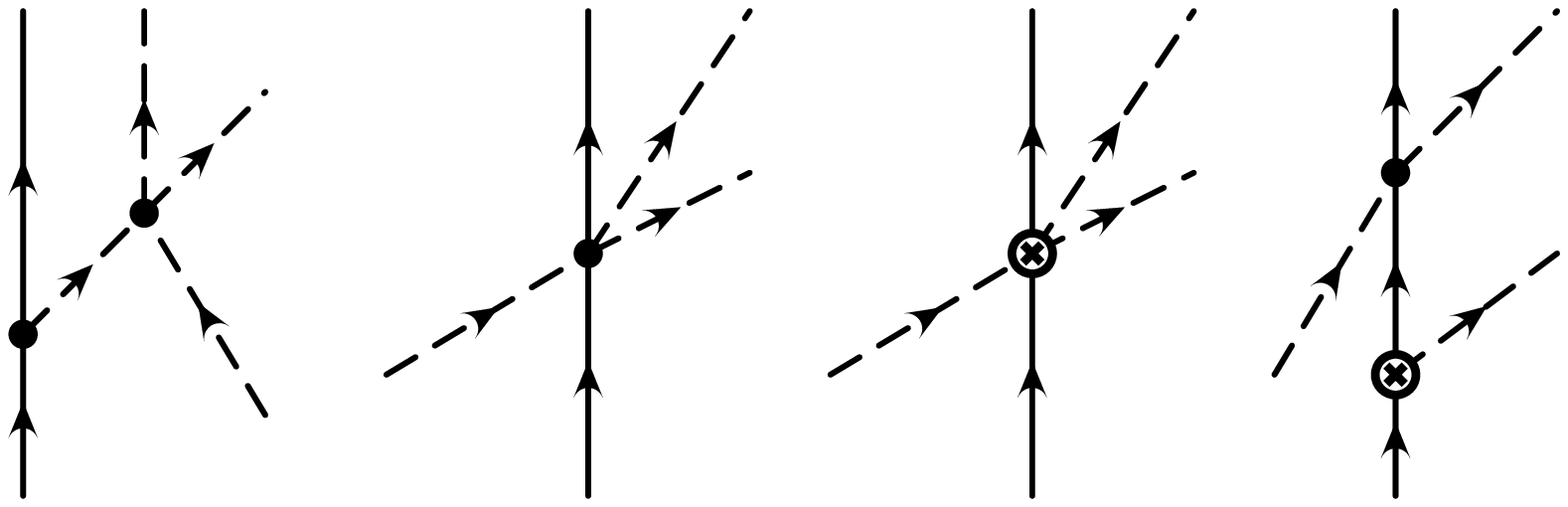}

\vspace{0.2cm}

\begin{center}

Fig.1: Diagrams which give the contributions to $D_{1,2}$
up--to--and--including order ${\cal O}(M_\pi)$. The circle--cross denotes an
insertion from ${\cal L}_{\pi N}^{(2)}$.

\end{center}

\vspace{0.3cm}

\noindent There are
potentially large contributions from diagrams with intermediate
$\Delta(1232)$ states of the type
$M_\pi^2 \, / \, ( \,m_\Delta - m - 2M_\pi \, )$,
which numerically would be of the order $10 \cdot M_\pi$.\footnote{Possible
large $\Delta$--contributions starting at order $M_\pi^2$ have yet to be
investigated in a systematic fashion together with loop effects and alike.}
We have checked
that no such terms appear from diagrams involving one or two
intermediate $\Delta$ resonances. Consequently, the chiral expansion is well
behaved but not too rapidly converging. The order $M_\pi$ corrections give
approximatively 50$\%$ of the leading term. As we will discuss below, the
calculations of Beringer \cite{beri} in relativistic baryon chiral perturbation
theory indicate that further $1/m$ suppressed kinematical corrections are
small. To get an idea about the corrections to eqs.(\ref{e7},\ref{e8}) we have
also calculated the imaginary parts of the threshold amplitudes from
the one--loop diagrams shown
in fig.2. These start to contribute at order $M_\pi^2$ with the result

\begin{equation}
{\rm Im} \, D_1  = -\frac{\sqrt{3} \, g_A^3 \, M_\pi^2}{128 \, \pi \, F_\pi^5}
 + {\cal O}(M_\pi^3) \label{e10}
\end{equation}

\begin{equation}
{\rm Im} \, D_2
= \frac{5 \, \sqrt{3} \, g_A^3 \, M_\pi^2}{64 \, \pi \, F_\pi^5}
 + {\cal O}(M_\pi^3) \label{e11}
\end{equation}

\noindent Of course, at this order there are other contributions
to the real parts of $D_{1,2}$, so one should
consider the $M_\pi^2$ corrections
given in eqs.(\ref{e10},\ref{e11}) as indicative.

\vspace{0.5cm}

\hspace{0.5in}
\epsfysize=1.5in
\epsfxsize=5in
\epsffile{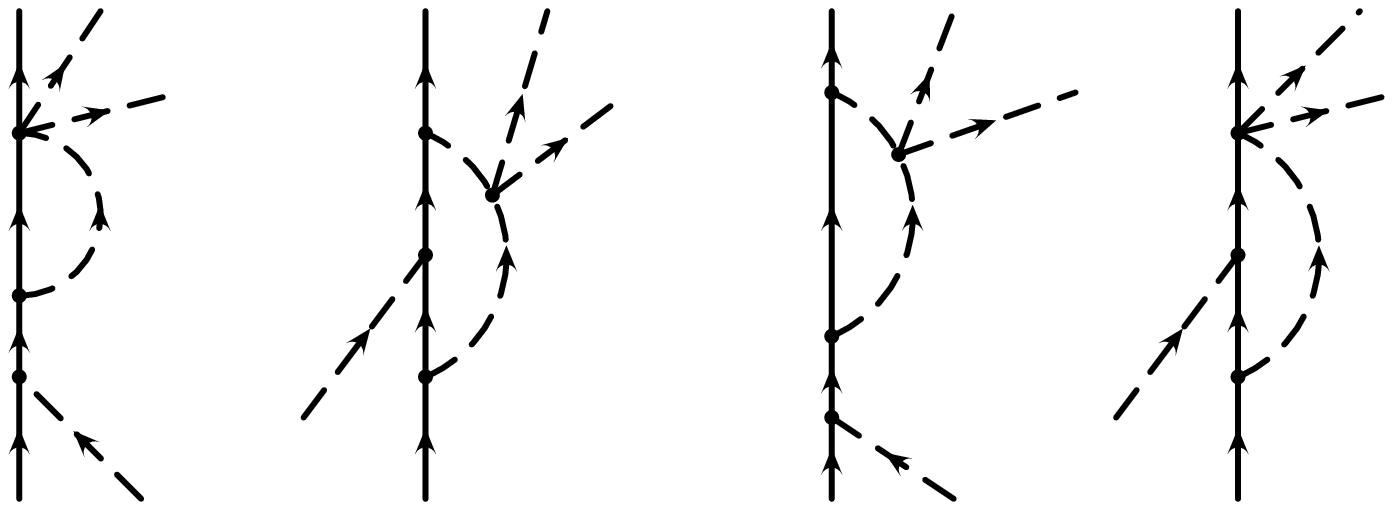}

\vspace{0.2cm}

\begin{center}

Fig.2: One--loop diagrams which give
a nonzero Im $D_{1,2}$ at order ${\cal O}(M_\pi^2)$.

\end{center}

\vspace{0.3in}

\noindent{\bf 4.} Let us now turn
to the numerical results. We use $F_\pi = 93$ MeV,
$g_{\pi N} = 13.4$, $m = 938.27$ MeV and $M_\pi = 139.57$ MeV. This amounts
to $D_1 = 2.4$ fm$^3$ and $D_2 = -6.8$ fm$^3$ or using eq.(\ref{e2})
\begin{equation}
{\cal A}_{32} \, = \, 2.4 \, M_\pi^{-3}, \quad {\cal A}_{10} \, = \,
5.5 \, M_\pi^{-3}
\label{e12}
\end{equation}
\noindent which compare favourably with the recent determinations of
ref.\cite{bulo}, ${\cal A}_{32} \, = \, 2.07 \pm 0.10 \, M_\pi^{-3}$ and
${\cal A}_{10} \, = \, 6.55 \pm 0.16 \, M_\pi^{-3}$. If one assumes that the
imaginary parts eqs.(\ref{e10},\ref{e11}) set the magnitude for the order
$M_\pi^2$ corrections of Re$D_{1,2}$, then one expects $D_1$ to change very
little and $D_2$ by approximatively 30$\%$.
 We have also calculated the cross sections for
 $\pi^+ p \to \pi^+ \pi^+ n$ and $\pi^- p \to \pi^0 \pi^0 n$ using
eq.(\ref{e3}). These are shown in fig.3 in comparison to the existing data.
Notice that we have calculated the matrix--elements in the isospin limit, in
fig.3 we have shifted the resulting cross sections  to account for
the corresponding thresholds as proposed by Beringer\cite{beri}.
 To a high degree of accuracy,
one can parametrize the cross sections calculated
from the first two terms of the chiral expansion of $D_{1,2}$ by the simple
forms using eq.(\ref{e4})

\begin{equation}
\sigma_{\rm tot}^{\pi^+ p \to \pi^+ \pi^+ n} (T_\pi)
  = 0.225 \, {\rm \mu b} \left(\frac{T_\pi - T_\pi^{\rm thr}}{10 \, {\rm MeV}}
\right)^2 \, \quad (T_\pi^{\rm thr} = 172.4 \, {\rm MeV})
\label{e13}  \end{equation}

\begin{equation}
\sigma_{\rm tot}^{\pi^- p \to \pi^0 \pi^0 n} (T_\pi)
  = 0.442 \, {\rm \mu b} \left(\frac{T_\pi - T_\pi^{\rm thr}}{10 \, {\rm MeV}}
\right)^2  \, \quad (T_\pi^{\rm thr} = 160.5 \, {\rm MeV})
\label{e14}
\end{equation}

\noindent as shown by the dashed lines in fig.3. The solid lines differ
very little
from the ones in ref.\cite{beri} indicating that higher order $1/m$ corrections
(which are summed up in the relativistic approach) are fairly small.
This also means that the approximation of using the exact threshold amplitude
in the threshold region is a very good one for the first 30 MeV. The
advantage of the heavy mass approach used here is the strict one--to--one
correspondence between the loop and small momentum expansion. The
abovementioned expectations of higher order corrections from Im$\, D_{1,2}$
are indeed such that they can improve the description of the data since the
first/second channel allows to test $D_1$/$D_2$, respectively.

\vspace{-0.5cm}

\hspace{1in}
\epsfysize=6in
\epsfxsize=4in
\epsffile{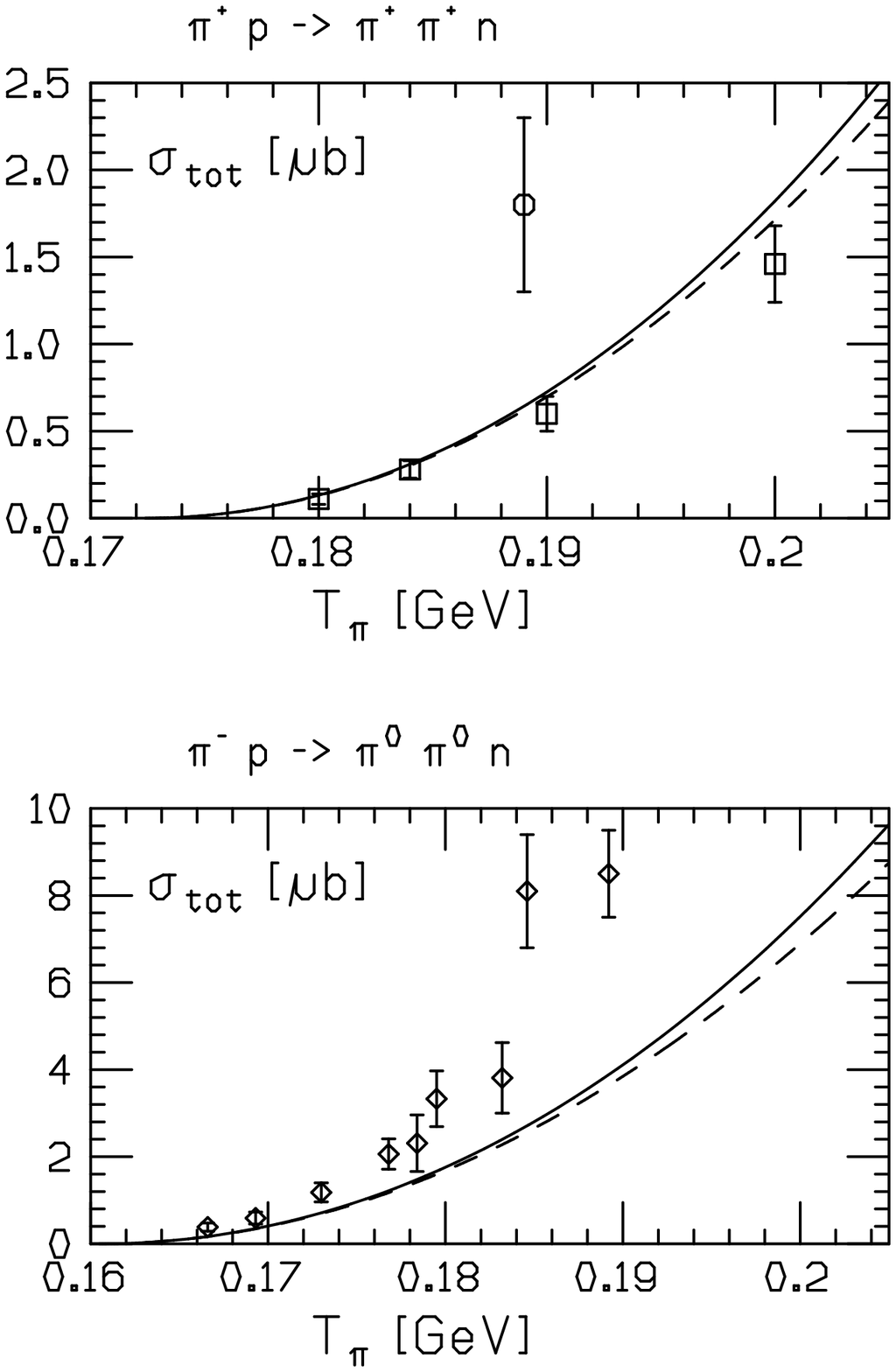}

\vspace{0.2cm}
\noindent Fig.3: Total cross sections for
$\pi^+ p \to \pi^+ \pi^+ n$ and $\pi^- p \to \pi^0 \pi^0 n$ in comparison to
the data. Squares: ref.\cite{sevi}, diamonds: ref.\cite{kern}
and octagon: ref.\cite{lowe}.

\vspace{0.5cm}

\noindent {\bf 5.} We have considered the first two coefficients
of the chiral expansion for the
threshold $\pi N \to \pi \pi N$ amplitudes
and derived a set of low--energy theorems,
eqs.(\ref{e7},\ref{e8},\ref{e10},\ref{e11}), which only
involve well--known physical parameters. We have also shown that
the corresponding
cross sections agree with the empirical ones close to threshold. Of course,
to go further, one has to consider loop diagrams as well as contributions
from resonance exchange (like e.g. the $\Delta (1232)$).
Ultimately, this will tell how
accurately one can in fact get to the elastic $\pi \pi$ amplitude from data on
single pion production. Work along these lines is in progress.

\vspace{0.5cm}

\noindent {\bf Acknowledgements}
\bigskip

\noindent We thank J. Stern for discussion and A. Schmidt for assistance.

\end{document}